\documentclass[aps,
twocolumn
,showpacs,preprintnumbers,amsmath,amssymb,superscriptaddress,prl,floatfix,10pt]{revtex4-2}

\usepackage{graphicx}
\usepackage{dcolumn}
\usepackage{bm}
\usepackage{graphics}
\usepackage{subfigure}
\usepackage{graphicx}
\usepackage{epsfig}
\usepackage{epstopdf}
\usepackage{xcolor}
\usepackage{multirow}
\usepackage{mathtools}
\usepackage{comment}
\usepackage{soul}
\usepackage[colorlinks=true,citecolor=blue]{hyperref}
\hypersetup{colorlinks=true,citecolor=blue,linkcolor=blue,urlcolor=blue}

\allowdisplaybreaks

\begin{document}
\title{Magnetotransport of tomographic electrons in a channel}

\author{Nitay Ben-Shachar}
\email{nshachar@caltech.edu}
\affiliation{School of Mathematics and Statistics, The University of Melbourne, Victoria 3010, Australia}
\affiliation{Department of Physics, California Institute of Technology, Pasadena CA, 91125, USA}

\author{Johannes Hofmann}
\email{johannes.hofmann@physics.gu.se}
\affiliation{Department of Physics, Gothenburg University, 41296 Gothenburg, Sweden}
\affiliation{Nordita, Stockholm University and KTH Royal Institute of Technology, 10691 Stockholm, Sweden}

\date{\today}

\begin{abstract}
Hydrodynamics is a new paradigm of electron transport in high-mobility devices, where frequent electron collisions give rise to a collective electron flow profile. However, conventional descriptions of these flows, which are based on the fluid equations for a classical gas extended to include impurity scattering, do not account for the distinct collisional relaxation in quantum-mechanical systems. In particular, by dint of Pauli blocking even modes of the distribution function relax over significantly shorter length scales than odd modes (dubbed the ``tomographic'' effect). We establish an analytical description of tomographic electron flow in a channel, and find four new distinguishing features: (i) Non-equilibrium effects from the boundaries penetrate significantly deeper into the flow domain; (ii) an additional velocity slip condition leads to a significant increase in the channel conductance; (iii) bulk rarefaction corrections decrease the curvature of the velocity profile in the channel center; and (iv) all these anomalous transport effects are rapidly suppressed with magnetic fields. The latter effect leads to a non-monotonic magneto-conductance, which can be used to measure both the even- and odd-mode mean free paths. 
Our asymptotic results unveil the underlying physics of tomographic flows and provide an alternative to numerical solutions of the Fermi-liquid equations.
\end{abstract}

\maketitle

\section{Introduction}

Electron transport is traditionally described in terms of the Drude picture, where charge carriers relax momentum by electron-phonon and electron-impurity scattering over a mean free path~$\ell_\text{MR}$, but they do not interact with each other~\cite{scaffidi2017,fritz24,baker24}. However, recent advances in experimental condensed matter physics have revealed electron flows that deviate from this description, and instead resemble the flow of a collisional classical gas~\citep{fritz24,varnavides23}. Such ``hydrodynamic'' flows have been observed in several high-mobility materials, including graphene~\citep{sulpizio19,bandurin16,palm24}, thin films of tungsten ditelluride~\citep{vool21}, and \mbox{PdCoO$_2$}~\citep{moll16}. Here, the electrons exhibit collective flow profiles observable in the form of current vortices~\cite{aharonsteinberg22} or Poiseuille flow through a channel~\cite{sulpizio19,ku20,vool21}. These hydrodynamic transport signatures are typically modeled using semiclassical descriptions---the Stokes-Ohm equation~\cite{fritz24,varnavides23} or its microscopic kinetic description based on a dual relaxation-time approximation~\cite{dejong95}---where an additional mean free path emerges for electron-electron scattering. Beyond the presence of impurity scattering, this is the same framework that would be used to describe the dynamics of a classical gas~\cite{knudsen09}, which prompts the immediate question: Is there anything that fundamentally distinguishes hydrodynamic electron flow from the flow of a classical gas?

\begin{figure*}[t]
    \centering
    \includegraphics{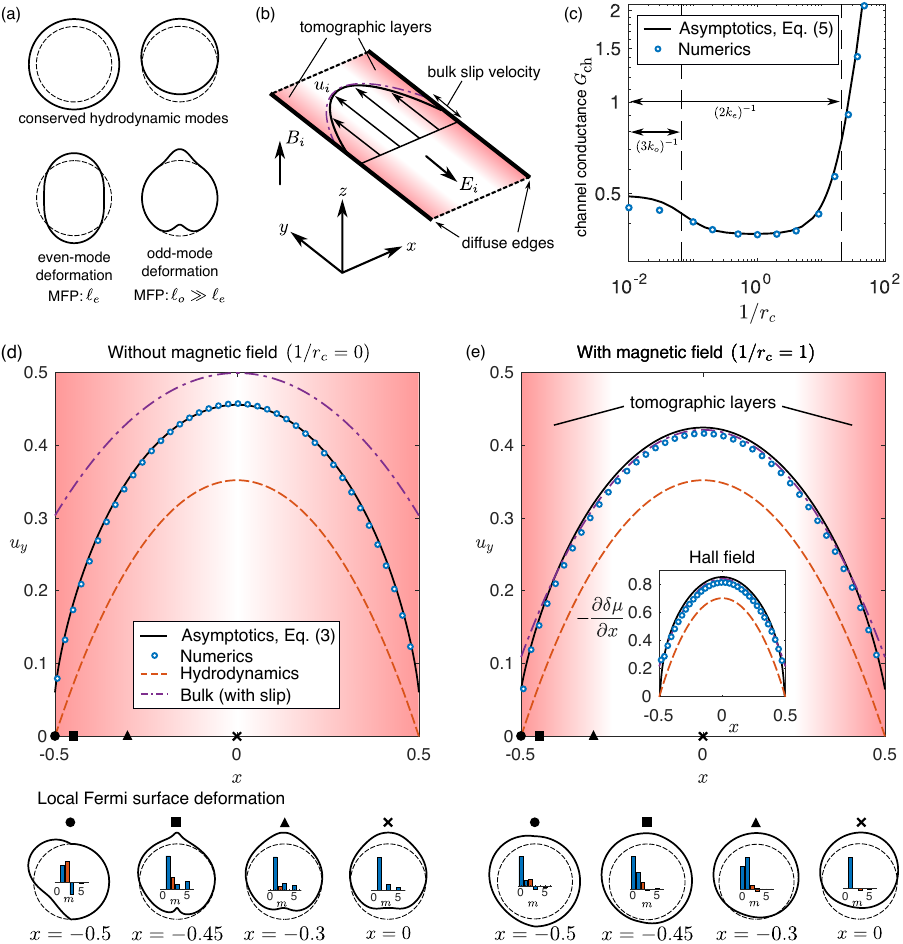}
    \caption{ 
    (a) Tomographic effect showing the relaxation mean-free path (MFP) of exaggerated deformations of the Fermi surface, with the global equilibrium denoted by a dashed line. (b) Sketch of flow in a channel as described in the main text. The approximate extent of the tomographic layers is indicated, and a sketch of the flow velocity profile (continuous black line), and that predicted in the bulk region (dash-dotted purple line). (c) Magnetoconductivity of a channel, scaled by the hydrodynamic conductivity \mbox{$e^2n\mathcal{E}L^3/(mv_F\ell_e)$}, with \mbox{$k_e=0.025,k_o=5,G\to\infty$} as obtained from direct numerical solutions to the Boltzmann equation (blue `$\circ$') and predicted by Eq.~\eqref{eq:conductance} (solid black line). The scale of the inverse cyclotron radius for the reduction (\mbox{$(3k_o)^{-1}$}) and enhancement (\mbox{$(2k_e)^{-1}$}) of the conductivity is indicated with dashed lines. (d) and (e): Flow in a channel driven by a constant electric field, with \mbox{$k_e=0.1$}, \mbox{$k_o=2$}, \mbox{$G=1$} and (d) \mbox{$1/r_c=0$}, (e) \mbox{$1/r_c=1$}. Velocity profile (scaled by \mbox{$e\mathcal{E}L^2/(mv_F\ell_e)$}) predicted by the present asymptotic theory (continuous black line), and previous hydrodynamic theory with no-slip boundary conditions (dashed orange line) and with slip conditions (dash-dotted purple line, see Eq.~\eqref{eq:ux_channel}) are compared with numerical solutions to the linearized Boltzmann equation (blue `$\circ$'). Shaded red regions indicate the approximate extent of the tomographic layers. The inset in (e) shows the Hall field, \mbox{$-\partial \delta \mu/\partial x$}. The exaggerated deformation of the Fermi surface described by the asymptotic solution are shown for \mbox{$x=0,-0.3,-0.45$} and \mbox{$-0.5$} (marked with `$\times$', `$\blacktriangle$', `$\blacksquare$' and `$\bullet$', respectively, in the channel for both (d) and (e)). Relative amplitude of the velocity modes \mbox{$\cos(m\theta)$} (blue) and \mbox{$\sin(m\theta)$} (red) are shown in the inset histograms. 
    }
    \label{fig:1}
\end{figure*}

This question is especially pertinent in view of recent literature studies~\citep{gurzhi95,ledwith17,ledwith19,hofmann23,nilsson24}, which have demonstrated that a second ballistic electronic mean free path emerges in addition to the short hydrodynamic mean free path for degenerate electrons: Only even modes of the electron distribution (deformations of the Fermi surface that are symmetric with respect to the electron velocity, illustrated in Fig.~\ref{fig:1}(a)) relax efficiently over a short hydrodynamic length scale \mbox{$\ell_e \ll L$}, where $L$ is the macroscopic length scale of the flow; microscopically, this relaxation is mediated by frequent head-on scattering~\cite{laikhtman92}. However, these scattering events do not relax the distribution function's odd-parity modes (deformations that are anti-symmetric with respect to the electron velocity, cf.~Fig.~\ref{fig:1}(a))~\cite{gurzhi95}. Odd-parity modes are thus anomalously long-lived with a significantly longer ballistic mean free path \mbox{$\ell_o \gtrsim L$}~\cite{gurzhi95,ledwith19,ledwith17,hofmann23,nilsson24}; this has been termed the ``tomographic effect''. Exact diagonalization studies of the electron collision integral show that the tomographic effect manifests in a low-temperature Fermi liquid (at temperatures \mbox{$T\lesssim0.1 T_F$}, where $T_F$ is the Fermi temperature)~\citep{hofmann23,nilsson24}. This effect has been shown to give rise to a wavelength-dependent viscosity~\citep{ledwith19} and a non-monotonic magneto-response~\citep{rostami24} in bulk flows. However, electron flows in realistic devices are sensitive to the scattering at device edges~\citep{palm24,aharonsteinberg22}, which are commonly modeled heuristically (e.g., using no-slip boundary conditions for the mean velocity at the boundary~\citep{ledwith19}) or included in numerical solutions~\citep{estrada24b,estrada24}, and a microscopic understanding of tomographic transport requires not only a description of bulk flow  but in finite device dimensions.

In this Letter, we study the electron flow in a channel with diffuse scattering at its edges [see Fig~\ref{fig:1}(b)], and establish from a systematic expansion of the Fermi-liquid kinetic equation that indeed the conventional semiclassical Stokes-Ohm or dual relaxation-time description of electron flow is fundamentally limited and does not describe electron flows in the degenerate regime. The theory we report reconciles this discrepancy and reveals the underlying governing physics for tomographic electron transport. In particular, we report an analytical description of four new distinguishing effects for tomographic flow in a channel: First, collisions establish a local equilibrium sufficiently far from the channel edges, where the flow is governed by Stokes-Ohm like equations with additional higher-derivative (rarefaction) corrections that must be included in a full description. These corrections decrease the curvature of the velocity profile compared to the hydrodynamic Poiseuille profile. Second, we show that scattering from device edges prohibits the establishment of a local equilibrium near the edges, creating a non-equilibrium layer that extends into the bulk over an anomalously large distance of \mbox{${\it O}(\sqrt{\ell_e \ell_o})$}, a region that we term the ``tomographic boundary layer''. Here, the electron system is characterized by a balance of diffusion mediated by even-mode scattering, ballistic odd-mode relaxation and the magnetic force. This layer is much larger than the conventional non-equilibrium boundary layer in classical hydrodynamics (known as the Knudsen layer), which extends over a much shorter distance~${\it O}(\ell_e)$~\cite{benshachar25a,benshachar25b,afanasiev22}. Third, the boundary conditions for the mean velocity in the bulk attain a significant slip at the boundaries of order~${\it O}(\ell_e \ell_o)$, which deviates from the slip-length boundary conditions that are widely used in the electron hydrodynamics literature~\cite{afanasiev22,palm24,pellegrino16,varnavides23,raichev22} and are ubiquitous in other fields of fluid dynamics~\citep{collis21,hadjiconstantinou21,sone07}. All of these effects establish that past hydrodynamic theories are inapplicable in describing the electron flows in recent experiments, which exhibit non-negligible even-mode mean-free path~$\ell_e$. Moreover, fourth, we show that as a characteristic signature of the tomographic transport regime, even a weak applied magnetic field strongly suppresses the tomographic layer and induces a transition to conventional magneto-hydrodynamic transport. We illustrate these phenomena for channel flow here, however, they are found to persist for an arbitrary device geometry, an analysis of which will be presented elsewhere~\cite{benshachar25LongPaper}.

An unequivocal macroscopic signature of the tomographic phenomena listed above is a non-monotonic channel magneto-conductance, from which both the odd and even-mode mean-free paths can be obtained [see Fig.~\ref{fig:1}(c)]. In particular, we find that even small magnetic fields suppress the additional tomographic velocity slip, which then decreases the channel conductance with increasing magnetic field, followed by an increase in the conductance at large fields due to the well-known suppression of the hydrodynamic viscosity. The suppression and subsequent increase manifest at cyclotron radii (the characteristic length scale of the magnetic field) comparable to~\mbox{$\ell_o$} and~\mbox{$\ell_e$}, respectively. Moreover, as we shall discuss, previously unreported tomographic phenomena interfere with past methodologies for inferring the Hall viscosity of electrons in a channel.

Our derivations are based on a solution of the  Fermi-liquid kinetic equation with diffuse boundary scattering at the channel edges. We go beyond the commonly used double relaxation-time approximation by including separate even and odd electron-electron relaxation mean-free-paths $\ell_e$ and $\ell_o$ in addition to momentum-relaxing collisions with mean-free-path $\ell_\text{MR}$, which satisfy 
\begin{align}
    \ell_e \ll L \lesssim \ell_o\ll \ell_\text{MR} . \label{eq:lengthscales}
\end{align}
In dimensionless form, we solve
\begin{equation}
	\begin{array}{rl} 
    &\displaystyle v_x \frac{\partial h}{\partial x} - 2 k_e v_y - \frac{1}{r_c} \varepsilon_{ij}v_j \frac{\partial h}{\partial v_i} \\[2ex]
    &= \displaystyle - \frac{1}{k_e} \left(\left[h\right]_e - \mu\right) - \frac{1}{k_o} \left(\left[h\right]_o - 2v_yu_y\right) - \frac{k_e}{G^2}\left(h - \mu\right) ,
    \end{array}
    \label{eq:linearBTE}
\end{equation}
where the flow is driven by a constant electric field in the negative $y$-direction, with an applied perpendicular magnetic field in the $z$-direction. Here, $[h]_{e(o)}$ indicates the even (odd) component of the distribution function $h$ at position $x$ across the channel, $v_i$ is the electron velocity and~\mbox{$r_c=R_c/L$} the scaled cyclotron radius, where~\mbox{$R_c=m^* v_F/(e\mathcal{B})$} is the dimensional cyclotron radius, $\mathcal{B}$ is the strength of the applied magnetic field, $m^*$ is the effective mass and $e$ is the fundamental charge. The spatial coordinate is scaled by the width of the channel $L$, and the electron velocity is scaled by the Fermi velocity~$v_F$. Scaled (i.e., dimensionless) variables are used henceforth. The first term on the right-hand side describes even-mode relaxation, where we define the dimensionless even-mode Knudsen number \mbox{$k_{e}= \ell_{e}/L$}. In the relaxation term, we exclude the density zero mode by subtracting the local perturbation to the electrochemical potential, \mbox{$\delta \mu = (2\pi)^{-1} \int_{-\pi}^\pi d\theta\, h$}, which is scaled by \mbox{$e\mathcal{E}L$} and with~\mbox{$\theta$} parameterizing the circular Fermi surface. The second term describes the separate odd-mode relaxation with odd-mode Knudsen number \mbox{$k_{o}= \ell_{o}/L$}, which does not relax the mean velocity, \mbox{$u_i = (2\pi)^{-1} \int_{-\pi}^\pi d\theta \, (v_i h)$}. The mean velocity has been scaled by the hydrodynamic velocity scale, \mbox{$e\mathcal{E}L^2/(m^*v_F\ell_e)$}, where $\mathcal{E}$ is the magnitude of the driving electric field. The Gurzhi number \mbox{$G=\sqrt{\ell_e \ell_\text{MR}}/L$} quantifies the strength of momentum-relaxing collision.

The length scale separation in Eq.~\eqref{eq:lengthscales} corresponds to \mbox{$k_e\ll 1$} and \mbox{$1\lesssim G,k_o$}. In the classical gas literature, a description of flows with small Knudsen numbers has been reported through analysis of the linearized Boltzmann equation using a matched asymptotic expansion~\citep{sone02,sone07}. This analysis techniques was recently employed to study conventional electron near-hydrodynamic flows~\citep{benshachar25a,benshachar25b}. Given the large odd-mode mean free path, however, the existence of an analogous expansion for tomographic flows is not obvious. The main advance of this work is a matched asymptotic expansion solution for tomographic flow in a channel, which indeed requires an analysis distinct to past studies of conventional near-hydrodynamic flows, evidenced by the non-analytic expansion parameter $\sqrt{k_e}$. Details of this expansion are reported in Methods, and
we summarize the results of this analysis next.

\section{Results}

\subsection{Velocity and Hall field profiles}

We begin by stating and describing the result for the velocity and Hall field profiles of electron flow through a channel with boundaries at \mbox{$x=\pm 1/2$} [cf. Fig~\ref{fig:1}(b)]. To linear order in $k_e$, we obtain the following scaled velocity profile across the channel
\begin{widetext}
\begin{equation}
        \begin{array}{rl} 
    &\displaystyle u_y =  G^2 \left(\frac{\cosh(\frac{2x}{G})}{\cosh(\frac{1}{G})}-1\right) + \biggl[4 k_ek_o \left(\mathcal{Y}_1\left(\frac{\frac{1}{2}+x}{\sqrt{k_ek_o}};\frac{k_o}{r_c}\right)+\mathcal{Y}_1\left(\frac{\frac{1}{2}-x}{\sqrt{k_ek_o}};\frac{k_o}{r_c}\right)\right) \\[1ex]
    &\quad \displaystyle - 2k_eG \tanh(\frac{1}{G}) \left(\mathcal{Y}_0\left(\frac{\frac{1}{2}+x}{\sqrt{k_ek_o}};\frac{k_o}{r_c}\right)+\mathcal{Y}_0\left(\frac{\frac{1}{2}-x}{\sqrt{k_ek_o}};\frac{k_o}{r_c}\right)\right) \biggr] + \biggl[k_e\left(\frac{k_o}{1+(3k_o/r_c)^2}+ \frac{64}{15\pi}G \tanh(\frac{1}{G})\right)\frac{\cosh(\frac{2x}{G})}{\cosh(\frac{1}{G})} \biggr] \\[1ex] 
    &\quad \displaystyle 
    + \biggl[\frac{k_ek_o}{1+(3k_o/r_c)^2} \frac{1}{2G\cosh(\frac{1}{G})}\Bigl(2x \sinh(\frac{2x}{G})-\cosh(\frac{2x}{G})\tanh(\frac{1}{G})\Bigr)\biggr]  + {\it O}(k_e^2) , 
        \end{array}
        \label{eq:ux_channel}
\end{equation}
and the Hall field profile
\begin{align}
    -\frac{\partial \delta \mu}{\partial x} = & \frac{2}{r_c}u_y -2 k_e \Bigg(\mathcal{T}_1\left(\frac{\tfrac{1}{2}+x}{\sqrt{k_ek_o}};\frac{k_o}{r_c}\right)+\mathcal{T}_1\left(\frac{\tfrac{1}{2}-x}{\sqrt{k_ek_o}};\frac{k_o}{r_c}\right)
     + \frac{G}{k_o} \tanh(\tfrac{1}{G}) \left[\mathcal{T}_0\left(\frac{\tfrac{1}{2}+x}{\sqrt{k_ek_o}};\frac{k_o}{r_c}\right)+\mathcal{T}_0\left(\frac{\tfrac{1}{2}-x}{\sqrt{k_ek_o}};\frac{k_o}{r_c}\right)\right] \Bigg) \nonumber \\&+  {\it O}(k_e^2),
    \label{eq:Ehall_channel}
\end{align}
\end{widetext}
which is the main result of this work, and is plotted in Fig.~\ref{fig:1}(d) and (e).

\begin{figure}[b!]
	\centering
    \includegraphics{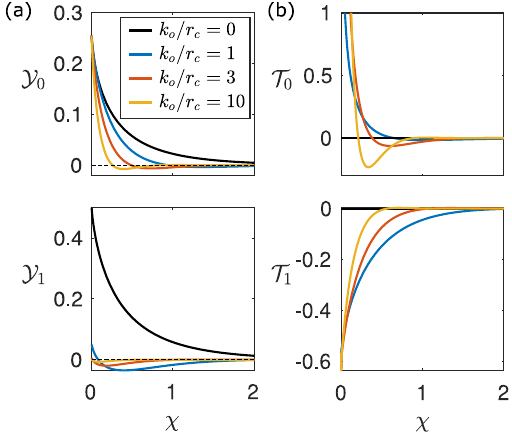}
    \caption{The tomographic layer functions $\mathcal{Y}_0,\mathcal{Y}_1, \mathcal{T}_0$ and $\mathcal{T}_1$, for \mbox{$k_o/r_c\in \{ 0,1,3,10\}$}, as indicated.}
	\label{fig:2}
\end{figure}

The first term in Eq.~\eqref{eq:ux_channel} [of order~\mbox{${\it O}(k_e^0)$}] is the well-known hydrodynamic solution that follows from the Stokes-Ohm equation with the no-slip boundary condition. The three subsequent terms in square brackets [of order~\mbox{${\it O}(k_e)$}] describe new phenomena discussed in the introduction: The first square bracket describes the tomographic layer correction, where $\mathcal{Y}_0$ and $\mathcal{Y}_1$---the ``tomographic boundary layer functions''---are shown in Fig.~\ref{fig:2}(a). These functions decay away from the boundaries and depend on a scaling variable~\mbox{$\chi=(1/2\pm x)/\sqrt{k_ek_o}$} that makes the extent of the layer over ${\it O}(\sqrt{k_ek_o})$ apparent. The second square bracket in Eq.~\eqref{eq:ux_channel} arises from the slip boundary conditions on the bulk equation
[see Eq.~\eqref{eq:slip} below]. This term is strictly positive and increases the velocity profile. The third square bracket is associated with finite-wavelength (rarefaction) corrections to the bulk equations at \mbox{${\it O}(k_ek_o)$}. Finally, a further hallmark of the tomographic transport regime is its rapid destruction by a magnetic field: The tomographic corrections in Eq.~\eqref{eq:ux_channel} (i.e., all terms of order \mbox{${\it O}(k_o)$}) decrease rapidly with increasing magnetic field (i.e., with decreasing cyclotron radius). This is seen directly from Eq.~\eqref{eq:ux_channel} for the slip condition and finite-wavelength corrections of \mbox{$O(k_ek_o)$}, which decrease in magnitude as a Lorentzian of width \mbox{$3k_o/r_c$}, while the boundary layer functions $\mathcal{Y}_{0/1}$ and $\mathcal{T}_{0/1}$ in Eqs.~\eqref{eq:ux_channel} and~\eqref{eq:Ehall_channel} become increasingly confined to the boundary. This is shown in Fig.~\ref{fig:2} for a dimensionless cyclotron radius \mbox{$k_o/r_c=1,3$}, and $10$ (blue, red, and yellow lines, respectively). 

Similarly, the Hall field in Eq.~\eqref{eq:Ehall_channel} at $O(k_e^0)$ [first term in Eq.~\eqref{eq:Ehall_channel}] is given by the Lorentz force as predicted by the continuum Stokes-Ohm equation~\citep{matthaiakakis2020}. However, at $O(k_e)$ [second term in Eq.~\eqref{eq:Ehall_channel}], non-continuum contributions arise in the tomographic layer, and are dictated by the functions $\mathcal{T}_0$ and $\mathcal{T}_1$, which are shown in Fig.~\ref{fig:2}(b). These tomographic layer functions vanish in the absence of a magnetic field (black lines), and the finite Hall response at finite magnetic field is confined closer to the boundary. In addition, the Hall field receives slip and rarefaction correction through the dependence of the velocity in the Lorentz term.

To illustrate these different contributions and to highlight the importance of a description beyond the Stokes-Ohm equation, Fig.~\ref{fig:1}(d) compares the velocity profile in Eq.~\eqref{eq:ux_channel} with parameters \mbox{$k_e=0.1$}, \mbox{$k_o=2$}, and \mbox{$G=1$} in the absence of an applied magnetic field at different levels of approximation. The full asymptotic solution~\eqref{eq:ux_channel} is strongly under-estimated by the hydrodynamic solution (i.e., the first term of Eq.~\eqref{eq:ux_channel} only) and strongly over-estimated by the bulk solution (i.e., when the tomographic layer functions $\mathcal{Y}_0$ and $\mathcal{Y}_1$ are omitted), respectively. Compared with the hydrodynamic prediction, the asymptotic solution exhibits a more rounded velocity profile near the center of the channel, which becomes significantly steeper near the channel edges. This shows that the solution beyond Stokes-Ohm derived here is required to obtain the correct shape of the velocity profile. While a hydrodynamic description requires the establishment of a local equilibrium, the phenomena discussed above are attributed to a significant deviation from local equilibrium throughout the channel (see Fermi surface deformations in Fig.~\ref{fig:1}(d)). This deviation is exemplified by the mode decomposition of the distribution function in Fig.~\ref{fig:1}(d), which is obtained by projecting the distribution function onto $\{\cos(m\theta),\sin(m\theta)\}$, for \mbox{$m\in \mathbb{Z}_{\geq 0}$}. The presence of \mbox{$m\geq 2$} modes indicates significant deviation from local equilibrium, which is characterized by the first two modes only (i.e., \mbox{$m=0,1$}, cf. Fig.~\ref{fig:1}(a)). The asymptotic solution~\eqref{eq:ux_channel} is also compared with a direct numerical solution of the linearized Boltzmann equation. Excellent agreement is observed, with similar agreement for other parameter values. The asymptotic solution~\eqref{eq:ux_channel} thus also provides an alternative to numerically expensive solutions of the Fermi liquid equations.

Since the odd-mode mean free path is a ballistic scale \mbox{$\ell_o \simeq L$}, even a moderate field (with a cyclotron radius \mbox{$R_c \simeq \ell_o$}) will suppress the tomographic phenomena discussed above, and the tomographic layer instead extends over the distance scale \mbox{${\it O}(\sqrt{\ell_e R_c})$}. We demonstrate this by plotting the velocity profile at various levels of approximation and in the presence of a magnetic field in Fig.~\ref{fig:1}(e), along with the induced Hall field. The velocity profile is observed to reduce in magnitude and become less rounded. Moreover, the discrepancy between the asymptotic solution (solid black line) and the bulk solution (dash-dotted purple line), which is dictated by the tomographic layer correction, reduces substantially. Thus, for \mbox{$R_c \lesssim \ell_o$} the flow transitions to a magneto-hydrodynamic form, and is well approximated by the bulk solution. This is further evidenced by the absence of sharp features in the Fermi surface deformations shown in Fig.~\ref{fig:1}(e).

\subsection{Channel conductance}

Integrating Eq.~\eqref{eq:ux_channel} over the channel width gives the conductance of the channel for \mbox{$k_e \lesssim 1$} at arbitrary magnetic field strength [Fig.~\ref{fig:1}(c)]. As is apparent from the figure, the enhancement of electron flow in the tomographic regime leads to an increased conductivity compared to the hydrodynamic solution. In addition, this additional boost to the electron current is suppressed  rapidly by small magnetic fields, i.e, the channel exhibits a negative magneto-conductance at weak magnetic fields up to \mbox{$R_c\sim \ell_o\gtrsim L$}. We quote here the result in the absence of bulk momentum relaxing collisions (\mbox{$G\to\infty$}),
\begin{equation}
    \begin{array}{rl}\displaystyle &G_\text{ch}=\displaystyle \int_{-1/2}^{1/2} dx \;u_y = \biggl(1+\biggl(\frac{2k_e}{r_c}\biggr)^2\biggr) \\[3ex]
    &\quad \displaystyle \times \left[\frac{1}{3}+k_e\frac{64}{15\pi}+\frac{k_ek_o}{1+(3k_o/r_c)^2}\right] + {\it o}(k_e) . \label{eq:conductance}
    \end{array}
\end{equation}
where we also include as a prefactor \mbox{$1+(2k_e/r_c)^2$} the positive magneto-conductance for strong magnetic fields, which  arises from the well-known reduction in the viscosity when \mbox{$R_c\sim \ell_e$}~\cite{alekseev16,scaffidi2017}. The first term in the square brackets is the hydrodynamic solution, while subsequent terms in the square brackets arise from the slip condition in Eq.~\eqref{eq:ux_channel}. This shows that for tomographic flows, the magneto-conductance of the channel is a non-monotonic function of the magnetic field strength, with a characteristic field dependence that allows to identify both the odd- and even-mode mean free paths. The conductance prediction in Eq.~\eqref{eq:conductance} is verified against numerical solutions of the linearized Boltzmann equation~\eqref{eq:linearBTE} in Fig.~\ref{fig:1}(c).

The results reported here complement qualitative expectations for channel flow given by~\citet{gurzhi95}. However, our resulting magneto-conductance differs from theirs (where it was suggested the conductance varies as \mbox{$R_c^{-1/3}$}). This is likely due to the qualitative nature of their discussion and the asymptotically larger odd-mode mean-free-path considered in their analysis. Furthermore,~\citet{ledwith19} reported a fractional scale-dependent conductivity for flow in a channel in the absence of a magnetic field when many modes participate in the bulk Fermi-surface deformation (i.e., due to finite-wavelength effects), and for odd-mode scattering rates which account for superdiffusion. However, this analysis did not account for diffuse reflection at boundaries, which were instead modeled by no-slip boundary conditions for the resulting velocity (current) profile~\citep{nazaryan24}.

\subsection{Hall viscosity}\label{sec:etaH_body}

As an application of our results, we proceed to investigate if past protocols to measure the Hall viscosity, which are based on the classical description of electron dynamics, also apply for tomographic electron flows. The Hall viscosity arises in the hydrodynamic equations describing electron flows in the presence of broken time-reversal symmetry (e.g., with an applied magnetic field)~\citep{scaffidi2017}, and is predicted to reduce the Hall field in a channel relative to its Lorentz force predictions~\citep{sulpizio19}. We consider two proposed methods to infer the Hall viscosity in view of tomographic flow: The first one, due to~\textcite{holder19}, employs spatial measurements of the Hall and current profiles in channel flow. The second, due to~\textcite{scaffidi2017},  utilizes the Hall resistance of channel flow. We show in the following that the former protocol is affected by bulk tomographic corrections to the governing equations, while the latter method (which utilizes the Hall resistance of flow in a channel) is affected by tomographic boundary layer corrections. These tomographic phenomena must be accounted for when inferring the Hall viscosity of tomographic flows.

\subsubsection{Spatial profile of the Hall field and current distribution}

\citet{holder19} propose to infer the Hall viscosity from spatial measurements of the Hall field and current distribution in the channel center (\mbox{$x=0$} in our notation). Within a hydrodynamic description, a Hall viscosity estimate~$\eta_H^{\rm sp}$ follows from the Hall field and velocity profile and their spatial derivatives at the center of the channel,
\begin{align}
    \frac{\eta_H^\text{sp}}{v_FL} = \frac{2}{r_c}\frac{1}{E_H''(0)}\left(E_H(0)-\frac{2}{r_c} u_y(0)\right) ,
    \label{eq:HolderHallVisc}
\end{align}
where~\mbox{$E_H=-\partial \delta \mu_B/\partial x$} is the scaled Hall field in the bulk, and we have rescaled the velocity profile compared to Ref.~\cite{holder19} to match our notation. This expression follows from an expansion of the Hall field near the center of the channel using the  kinetic equation. However, substituting the bulk governing equations for tomographic flow [see Eq.~\eqref{eq:bulkEqs} below] into Eq.~\eqref{eq:HolderHallVisc} gives
\begin{align}
    \frac{\eta_H^\text{sp}}{v_FL} = \frac{k_e^2}{2r_c}+\frac{k_e^2}{2r_c} \times \frac{3k_o^2}{2G^2}\frac{1}{1+(3k_o/r_c)^2} .
\end{align}
The first term is indeed associated with the usual hydrodynamic Hall viscosity, consistent with Ref.~\cite{holder19}, but there is now an additional  second term that follows from finite-wavelength tomographic correction. Again, this correction is suppressed at strong magnetic fields compared to the hydrodynamic result. It also introduces a dependence of $\eta_H^\text{sp}$ on the disorder strength, and it vanishes for clean systems where \mbox{$G\to\infty$}. When extracting the hydrodynamic Hall viscosity from experimental measurements of the Hall field, this tomographic finite-wavelength such a correction must be taken into account.

\subsubsection{Hall resistance}

\citet{scaffidi2017} propose to infer the Hall viscosity by measuring the deviation of the Hall resistance across a channel from its continuum (Ohmic) value. This dimensionless channel Hall resistance, which is scaled by~\mbox{$\mathcal{B}/(ne)$}, is given by
\begin{align}
    R_{xy} = \frac{r_c}{2} \frac{\left.\delta \mu\right|_{x=-1/2}-\left.\delta \mu\right|_{x=1/2}}{\int_{-1/2}^{1/2} dx \, u_y}. \label{eq:Rxy}
\end{align}
Within a hydrodynamic description (i.e., using the Stokes-Ohm equation) and omitting the Hall viscosity contribution, this quantity is equal to unity,~\mbox{$R_{xy}=1$}. However, including the Hall viscosity, $R_{xy}$ will be reduced from this bulk value. In the absence of bulk momentum-relaxation (i.e., for~\mbox{$G\to \infty$}), the deviation is predicted as~\citep{scaffidi2017}
\begin{align}
    \Delta R_{xy}^H = R_{xy}^H-1 = -\frac{6k_e^2}{1+(2k_e/r_c)^2} \propto \eta_H, \label{eq:DeltaRxyH}
\end{align}
where superscript `H' indicates the past hydrodynamic prediction. Equation~\eqref{eq:DeltaRxyH} is derived using the well-known relation between the velocity and Hall profiles of hydrodynamic flow~\citep{scaffidi2017,sulpizio19},
\begin{align}
    - \frac{\partial \delta \mu^H}{\partial x} = \frac{2}{r_c}\left(u^H_y + \frac{k_e^2}{2(1+(2k_e/r_c)^2)}\frac{\partial^2 u_{y}^H}{\partial x^2}\right), \label{eq:channl_EHall_eqn}
\end{align}
where the first term in the brackets above is the Lorentz force (i.e., the usual Hall field), and the second term is the contribution of the Hall viscosity, which reduces the Hall field at \mbox{${\it O}(k_e^2)$}.

Interestingly, this proportionality between $\Delta R_{xy}$ and $\eta_H$ in Eq.~\eqref{eq:DeltaRxyH} is obscured for tomographic electron flow. Instead, we find that to leading-order $\Delta R_{xy}$ is set by the tomographic boundary layer: Substituting the asymptotic solution in Eqs.~\eqref{eq:ux_channel} and~\eqref{eq:Ehall_channel} into Eq.~\eqref{eq:Rxy} gives
\begin{align}
        &\Delta R_{xy}=R_{xy}-1 = k_e^{3/2} \frac{2r_c k_o^{1/2}}{G^2(G\tanh(\frac{1}{G})-1)} \nonumber \\
        &\times \int_0^\infty d\chi \, \Bigl[\frac{G \tanh(\frac{1}{G})}{k_o} \mathcal{T}_0\Bigl(\chi;\frac{k_o}{r_c}\Bigr) - \mathcal{T}_1\Bigl(\chi;\frac{k_o}{r_c}\Bigr)\Bigr] .  
        \label{eq:DeltaRxyTomo}
\end{align}
Equation~\eqref{eq:DeltaRxyTomo} reveals that \mbox{$\Delta R_{xy}$} is not proportional to the Hall viscosity (which is $O(k_e^2)$, cf. Eq.~\eqref{eq:DeltaRxyH}), and is instead dominated by $O(k_e^{3/2})$ contributions from the tomographic layer (i.e., the functions $\mathcal{T}_0$ and $\mathcal{T}_1$). Importantly, these tomographic layer contributions to $\Delta R_{xy}$ are asymptotically larger than the contribution from the Hall viscosity. 
In fact, substituting our asymptotic predictions in Eqs.~\eqref{eq:ux_channel} and~\eqref{eq:Ehall_channel} into Eq.~\eqref{eq:Rxy} for \mbox{$k_e=0.025, k_o=5, G\to\infty$} and \mbox{$1/r_c=0.02$}, gives \mbox{$\Delta R_{xy}=+0.20$}, in excellent agreement with direct numerical solutions to the linearized Boltzmann equation which give \mbox{$\Delta R_{xy}^\text{num}=+0.19$}. However, this result is of opposite sign to the prediction using the Hall viscosity alone (see Eq.~\eqref{eq:DeltaRxyH}). Hence, when tomographic effects are present, it appears that the value of \mbox{$\Delta R_{xy}$} does not provide an accurate avenue from which to infer the Hall viscosity, which is instead dominated by near-boundary kinetic effects.

\section{Discussion}

The results presented above demonstrate that a conventional semiclassical description of hydrodynamic electron flow in terms of the Stokes-Ohm equation is inapplicable at low temperatures where Pauli blocking introduces a ballistic odd-parity mean free path in addition to a short hydrodynamic scale. The ensuing tomographic flow is very different from hydrodynamic flow and marked by a large kinetic boundary layer, which reflects the mixed ballistic and hydrodynamic relaxation, a more rounded velocity profile, and enhanced conductance. Furthermore, a direct hallmark of tomographic flow is the rapid suppression of these phenomena by even weak magnetic fields. In particular, combining this new prediction with the known reduction of the hydrodynamic viscosity with magnetic field shows a new minimum in a channel magneto-conductance, which can be used to infer both the even- and odd-mode electronic mean-free paths. Beyond channel flow, our results indicate that the appropriate starting point of a hydrodynamic description of electrons at low temperatures should be the generalized Boltzmann equation~\eqref{eq:linearBTE} and not the Stokes-Ohm or a dual-relaxation time description. 

In current experiments in ultra-clean materials, the hydrodynamic electron transport regime is realized in an intermediate temperature regime below the Fermi temperature~\mbox{$T\lesssim T_F$}. Here, with decreasing temperature, Pauli blocking increases the electron mean free path compared to a classical gas, with an even stronger  increase of the momentum-relaxing mean free path by phonon scattering, inducing a crossover from standard Drude to hydrodynamic transport. The corresponding odd- and even-parity mean free paths have comparable magnitude as dictated by standard Fermi liquid scaling~\cite{nilsson24}, and are small compared to both the device dimension and the momentum-relaxing mean free path, i.e., \mbox{$\ell_e \simeq \ell_0 \sim (T_F/T)^2 \ll l_{\rm MR}, L$}. The crossover from the hydrodynamic to tomographic electron transport regime is then expected at lower temperatures \mbox{$T\lesssim 0.1 T_F$}~\cite{nilsson24}, where the odd-parity mode become anomalously long-lived compared to the Fermi liquid scaling, with an asymptotic scaling  \mbox{$\ell_e \sim (T_F/T)^2$} and \mbox{$\ell_o\sim(T_F/T)^4$} at very low temperatures~\cite{ledwith19,hofmann23,nilsson24}. Such a temperature range is readily accessible for electron Fermi liquids, and the tomographic regime which satisfies the length scale separation in Eq.~\eqref{eq:lengthscales} is experimentally viable provided that devices are sufficiently clean that impurity scattering does not dominate the electronic mean free path. Our description now captures the full hydrodynamic to tomographic crossover in channel transport in closed analytical form with only two adjustable parameters in the form of the odd- and even-parity mean free paths. Besides the prediction for the overall magneto-conductance, the signatures we discuss should be readily detectable in present-day experiments that allow a direct position-resolved measurement of both the velocity and Hall field profiles~\citep{sulpizio19,palm24}.

\section{Methods}

Equations~\eqref{eq:ux_channel},~\eqref{eq:Ehall_channel} and~\eqref{eq:conductance} are obtained from an asymptotic solution to the Fermi liquid kinetic equation~\eqref{eq:linearBTE} for \mbox{$k_e\ll 1$}. Diffuse scattering from the device boundaries at \mbox{$x=\pm1/2$} is assumed, which dictates the reflected distribution function as
\begin{equation}
    h \bigr|_{x=\pm 1/2} = -\frac{1}{2} \int_{v_x\lessgtr0} d\theta \, (h v_x) , \quad v_x\gtrless 0 , \label{eq:diffuse}
\end{equation}
where the upper (lower) inequality sign holds for the left (right) boundary. 
This boundary condition is illustrated in the Fermi surface deformations at the left boundary shown in Figs.~\ref{fig:1}(d) and (e). 
The distribution function is expressed as a sum of a bulk region solution and a tomographic layer correction \mbox{$h=h_B+h_T$}, respectively denoted with subscripts ``B'' and ``T''. The latter term (i.e., the tomographic layer correction) is appreciable for \mbox{$|x\pm1/2|\sim \sqrt{k_ek_o}$} only and decays away from the device edges. Each of the moments is expressed with a similar sum.

Substituting this sum into the kinetic equation~\eqref{eq:linearBTE} gives governing equations for $h_B$ and $h_T$. Each of $h_B$ and $h_T$ is expressed as a regular perturbation expansion in $\sqrt{k_e}$,
\begin{equation}
    h_\alpha = h_\alpha^{(0)} + \sqrt{k_e} h_\alpha^{(1)} + k_e h_\alpha^{(2)} + \ldots \label{eq:keExpansion}
\end{equation}
for \mbox{$\alpha\in\{B,T\}$} and likewise for the moments $\delta\mu$ and $u_i$. Substituting this expansion into Eq.~\eqref{eq:linearBTE} and collecting powers of $k_e$ gives a set of coupled equations for each $h_B^{(n)}$, which are solved sequentially. Computing the zeroth and first moments of these solutions then gives a set of bulk governing equations for the macroscopic variables $\delta \mu_B^{(n)}$ and $u_{B|y}^{(n)}$, which are functions of $x$ only. These are (up to \mbox{$n=2$} for \mbox{$u_{B|y}^{(n)}$} and \mbox{$n=4$} for \mbox{$\delta \mu_B^{(n)}$})
\begin{equation}
    \begin{array}{rl} \displaystyle \frac{1}{2}\frac{\partial \delta \mu^{(n)}_B}{\partial x} + \frac{1}{r_c} u_{B|y}^{(n)} =& \displaystyle \mathcal{J}_n, \\[2ex] \displaystyle 
    -\dfrac{1}{4}\dfrac{\partial^2 u_{B|y}^{(n)}}{\partial x^2}+\dfrac{u_{B|y}^{(n)}}{G^2}=& \displaystyle \mathcal{I}_n,
    \end{array}\label{eq:bulkEqs}%
\end{equation}
with
\begin{subequations}
\begin{align}
    \mathcal{J}_0=&\mathcal{J}_1=\mathcal{J}_2=\mathcal{J}_3=0, \\ 
    \mathcal{J}_4 = & \frac{1}{2r_c}\frac{\partial^2 u_{B|y}^{(0)}}{\partial x^2} \left[1+\frac{3k_o^2}{2G^2 (1+(3k_o/r_c)^2)}\right] \\
    \mathcal{I}_0 = & \mathcal{I}_1=0, \quad \mathcal{I}_2 = \dfrac{k_o}{16}\frac{1}{1+(3k_o/r_c)^2} \dfrac{\partial^4 u_{B|y}^{(0)}}{\partial x^4} .\label{eq:calJI}
\end{align}
\end{subequations}
The widely-used incompressible Stokes-Ohm equations are retrieved at leading-order in $k_e$ (i.e., \mbox{$n=0$}), reflecting that a local equilibrium is established for \mbox{$k_e\to 0$} despite the weak odd-mode relaxation (i.e., finite $k_o$). However, at \mbox{$O(k_e)$} (i.e., order \mbox{$n=2$}), previously unreported terms arise that are associated with finite wave-length effects and the tomographic collision operator, where the last term ${\cal I}_2$ in Eq.~\eqref{eq:bulkEqs} gives rise to the finite-wavelength correction in Eq.~\eqref{eq:ux_channel}. Moreover, the Hall field deviates from that dictated by past near-hydrodynamic theories at \mbox{$O(k_e^2)$} (i.e., \mbox{$n=4$}) due to tomographic phenomena. In particular, the first term in square brackets in $\mathcal{J}_4$ (i.e., the `1' in Eq.~\eqref{eq:calJI}) is the usual Hall viscosity contribution, which corresponds to the well-known dimensional Hall viscosity for \mbox{$R_c\gg \ell_e$} of \mbox{$\eta_H=v_F \ell_e^2/(2R_c)$}. However, the second term in square brackets in \mbox{$\mathcal{J}_4$} (i.e., the term of \mbox{$O(k_o^2)$} in Eq.~\eqref{eq:calJI}) is a previously unreported correction to the Hall field that competes with the Hall viscosity in dictating the Hall field.

The bulk distribution functions that follow from Eq.~\eqref{eq:bulkEqs} can only satisfy the diffuse boundary condition~\eqref{eq:diffuse} at \mbox{${\it O}(1)$} and \mbox{${\it O}(\sqrt{k_e})$}, which gives a no-slip boundary condition for the bulk equations at these orders, and zero tomographic layer corrections (i.e., \mbox{$h_T^{(0)}=h_T^{(1)}=0$}). A tomographic layer correction must be added to the distribution function at \mbox{${\it O}(k_e)$} to satisfy Eq.~\eqref{eq:diffuse}. Expanding $h_T$ in powers of $\sqrt{k_e}$ (see Eq.~\eqref{eq:keExpansion}) shows that $h_T^{(2)}$ is an odd function that satisfies the ``tomographic equation''
\begin{equation}
    \begin{array}{rl}&\displaystyle v_x^2 \frac{\partial^2 h_{T}^{(2)}}{\partial \chi^2}-h_T^{(2)} +\frac{k_o}{r_c} \varepsilon_{ij}v_j \frac{\partial h_T^{(2)}}{\partial v_i} \\[2ex] &\displaystyle \hspace{2em}= \mp v_x\sqrt{k_o}\,\frac{\partial \delta\mu_T^{(3)}}{\partial \chi} - 2v_yu_{T|y}^{(2)}, \end{array}\label{eq:govEqnhT2}
\end{equation}
written here with a rescaled boundary-normal coordinate \mbox{$\chi = (1/2\pm x)/\sqrt{k_ek_o}$}, where the top and bottom signs are taken for the tomographic layer near \mbox{$x=-1/2$} and \mbox{$x=1/2$}, respectively. Equation~\eqref{eq:govEqnhT2} possesses an identical structure to the governing tomographic equation of~\citet{ledwith19} (Eq.~(10) of~\citep{ledwith19} with \mbox{$p=0$}). Solution for $h_T^{(2)}$ gives (i) boundary conditions for the bulk equations at ${\it O}(k_e)$, and (ii) tomographic layer corrections to the distribution function and macroscopic variables at this order. Multiplying Eq.~\eqref{eq:govEqnhT2} by $v_y$ and integrating over the electron velocity gives \mbox{$\int_{-\pi}^\pi d\theta (v_y-v_y^3) h_T^{(2)} =0$}, which, in conjunction with Eq.~\eqref{eq:diffuse}, gives the velocity slip boundary conditions for the bulk equations at ${\it O}(k_e)$,
\begin{equation}
    u_{B|y}^{(2)} = \pm \frac{32}{15\pi}\frac{\partial u_{B|y}^{(0)}}{\partial x} - \frac{k_o}{4(1+(3k_o/r_c)^2)} \frac{\partial^2 u_{B|y}^{(0)}}{\partial x^2} .\label{eq:slip}
\end{equation}
The solution of the bulk equations in Eq.~\eqref{eq:bulkEqs} at this order with these slip boundary conditions gives the second term in brackets in Eq.~\eqref{eq:ux_channel}. Interestingly, the slip condition in Eq.~\eqref{eq:slip} takes a form identical to the ``second-order'' slip condition used in the modeling of a slightly rarefied gas~\citep{hadjiconstantinou03,deissler64}. Finally, the tomographic layer corrections to the velocity and Hall field is expressed in a form identical to Eq.~\eqref{eq:slip},
\begin{subequations}
\begin{align}
    u_{T|y}^{(2)} =& \mp \frac{\partial u_{B|y}^{(0)}}{\partial x}\mathcal{Y}_0(\chi;k_o/r_c) + \frac{k_o}{2} \frac{\partial^2 u_{B|y}^{(0)}}{\partial x^2} \mathcal{Y}_1(\chi;k_o/r_c), \label{eq:uT} \\
    \frac{\partial \delta \mu_{T}^{(3)}}{\partial \chi} =&  \mp \frac{\partial u_{B|y}^{(0)}}{\partial x}\mathcal{T}_0(\chi;k_o/r_c) + \frac{k_o}{2} \frac{\partial^2 u_{B|y}^{(0)}}{\partial x^2} \mathcal{T}_1(\chi;k_o/r_c).
\end{align}
\end{subequations}
Substituting these into Eq.~\eqref{eq:govEqnhT2} and solving the resulting equation numerically gives the tomographic layer functions in Fig.~\ref{fig:2}, and hence the first term in square brackets in Eq.~\eqref{eq:ux_channel}.

\begin{acknowledgments}
This work is supported by Vetenskapsr\aa det (Grant Nos. 2020-04239 and 2024-04485), the Olle Engkvist Foundation  (Grant No. 233-0339), the Knut and Alice Wallenberg Foundation, and Nordita. We acknowledge support from The University of Melbourne’s Research Computing Services and the Petascale Campus Initiative. N.B.S. acknowledges support from the Australian Government Research Training Program Scholarship.
\end{acknowledgments}

\bibliography{letter}

\end{document}